**Main Manuscript for**

# Tunable vortex Majorana modes controlled by strain in homogeneous LiFeAs

Wenyao Liu[1,2†,#], Quanxin Hu[1,2†], Xiancheng Wang[1†], Yigui Zhong[3], Fazhi Yang[1,2], Lingyuan Kong[1,##], Lu Cao[1,2], Geng Li[1,2,4,5], Yi Peng[1,2], Kozo Okazaki[3,6,7],Takeshi Kondo[3,7], Changqing Jin[1,2,4], Jinpeng Xu[1,2,5*], Hong-Jun Gao[1,2,5*], and Hong Ding[1,5,8*]

[1]Beijing National Laboratory for Condensed Matter Physics and Institute of Physics, Chinese Academy of Sciences; Beijing 100190, China

[2]School of Physical Sciences, University of Chinese Academy of Sciences, Beijing 100190, China

[3]Institute for Solid State Physics, University of Tokyo, Kashiwa, Chiba 277-8581, Japan

[4]Songshan Lake Materials Laboratory, Dongguan, Guangdong 523808, China

[5]CAS Center for Excellence in Topological Quantum Computation, University of Chinese Academy of Sciences, Beijing 100190, China

[6]Material Innovation Research Center, The University of Tokyo, Kashiwa, Chiba 277-8561, Japan

[7]Trans-scale Quantum Science Institute, The University of Tokyo, Bunkyo-ku, Tokyo 113-0033, Japan

[8]Tsung-Dao Lee Institute, Shanghai Jiao Tong University, Shanghai 201210, China

[†]These authors contributed equally to this work

[#]Present Address: Laboratory for Assembly and Spectroscopy of Emergence, Boston College, MA, USA

[##]Present Address: T. J. Watson Laboratory of Applied Physics & Institute for Quantum Information and Matter, California Institute of Technology, Pasadena, CA, USA

[*]Correspondence to: dingh@sjtu.edu.cn, hjgao@iphy.ac.cn, xujp@iphy.ac.cn

**This PDF file includes:**

Main Text
Figures 1 to 4

## Abstract

The iron-based superconductors (FeSCs) have recently emerged as a promising single-material Majorana platform by hosting isolated Majorana zero modes (MZMs) at relatively high temperatures. To further verify its Majorana nature and move forward to build topological quantum qubits, it is highly desirable to achieve tunability for MZMs on homogeneous FeSCs. Here, with an in-situ strain device, we can controllably create MZMs on the homogeneous surface of stoichiometric superconductor LiFeAs by altering its chemical potential. The evolution of discrete energy modes inside a strained vortex is found to mimic exactly as the predicted topological vortex case, proving the Majorana nature of emerging zero modes of vortex. More importantly, our work provides a controllable method for MZM in a homogeneous FeSC, and such achievement of tunability of MZMs in the FeSC Majorana-material platform is an important step towards their application in topological quantum computation.

## Significance Statement

Our work provided the direct evidences between the vortex zero modes and superconducting topological surface state in LiFeAs by switching the vortex phase through the in-suit strain device, which gives perhaps the strongest evidence for Majorana zero modes in FeSCs. More importantly, our achievement proves that the uniaxial strain is an effective method to artificially adjust the chemical potential and therefore control the appearance of vortex MZM, and our result should enlighten the applicable design of FeSCs device for the future MZM-braiding experiment.

## Main Text
## Introduction

Majorana zero modes (MZMs), which have potential applications for fault-tolerant quantum computation due to their non-Abelian statistics, have attracted intensive interest in recent years (*1-2*). Compared with other predicted proposals for MZMs, *e.g.,* intrinsic *p*-wave superconductors and heterostructures with strong spin-orbital coupling and superconductivity (*3-13*), the topological-nontrivial iron-based superconductors (FeSCs) have recently emerged as a promising platform in study and application of MZMs, with unique advantages including higher transition temperatures, larger topological gaps and simpler material synthesis (*14-27*). However, other technical obstacles, like inhomogeneity of the topological-phase region in materials and lack of applicable controlling methods of MZMs, remain as roadblocks to MZM braiding for this new platform (*28, 29*). In addition, more direct and solid evidence is needed to declare the existence of isolated MZM with full confidence. Motived by these two goals, we recently discovered that the stoichiometric FeSC material, LiFeAs (*30*), could provide a good opportunity to overcome these obstacles. In this work, we use the uniaxial strain as a tuning knob to alter the chemical potential ($\mu$) of LiFeAs, which can change the phase of vortices from "trivial vortex region" to "Kitaev vortex region" (31,32), and consequently switch on-and-off the MZMs inside the magnetic vortices in this homogeneous superconductor.

## Results

### Topological band structure in LiFeAs

LiFeAs has a dopant-free stoichiometric crystal structure (Fig. 1A) and a charge-neutral lithium-atom terminating surface after cleavage (33-35), where most of the exposed area is a uniform ordered tetragonal lattice interspersed with a sparse distribution of defects (Fig. 1B). As a recent angle-resolved photoemission spectroscopy (ARPES) work reported (*21*), multiple topological bands exist in LiFeAs (Fig. 1C), including the topological insulator (TI) surface band and the bulk Dirac fermion of topological Dirac semimetal (TDS) phase. Theoretical works predicted (36,37) that either TI or TDS phase could lead to Majorana quasiparticles as localized single zero modes or mobile helical modes inside vortex cores. Therefore, LiFeAs seems to be a good candidate of a homogenous Majorana platform (*33,35*). However quite puzzlingly, no zero mode has been observed in the vortices of LiFeAs in the previous scanning tunneling microscopy/spectroscopy (STM/S) work (33).

The explanation for this apparent paradox likely lies on the inopportune location of the chemical potential relative to the Dirac point of the TI state in LiFeAs. A recent hypothesis has proposed that the Fermi energy ($E_F$) of LiFeAs at the impurity-free region is located right on the bending region of the Dirac surface state (the yellow shaded region in Fig. 1D). In this case(35), when $E_F$ crosses the helical Dirac bands twice (Fig. 1D), two MZMs will emerge simultaneously in a single vortex core (Fig. 1F), which leads to the annihilation between these two MZMs. Therefore, the material is still like locating at the "trivial vortex region", although $E_F$ has likely crossed the TI surface state. An alternative scenario for absence of MZMs is that the Fermi level of the material in fact lies above the TI surface state (indicated as "$E_F'$" in Fig. 1D), which means the pristine LiFeAs should be in the "trivial vortex region" as suggested by theoretical works (31,32), where the topological surface state does not participate in forming the superconducting vortex. Consequently, although the material itself is still topological nontrivial, in an impurity-free vortex of LiFeAs, no zero-bias conductance peak (ZBCP) exists, and the non-zero bound states mainly originate from the bulk bands (Fig. 1E), as reported by previous STM/S studies (33-35, 38).

Fortunately, this inopportune position of $E_F$ can be tuned away if one could adjust the chemical potential. For instance, some impurities/disorders can alter the chemical potential locally (35), and thus MZMs can be found in specific impurity-assisted vortices. However, impurities or disorders are seldom controllable, and their presence may also poison the MZMs, making braiding of MZMs difficult or even impossible. Therefore, we need a more controllable way to alter the position of $E_F$, especially if we could lower $E_F$ towards the TI Dirac cone. Below we will demonstrate that the uniaxial strain is such an ideal method.

### Evolution of band dispersion under uniaxial strain

We recently realized that applying a uniaxial strain on LiFeAs could shift its topological bands as required. Consequently, we carried out high-resolution ARPES measurements on LiFeAs with a home-made uniaxial-strain device. The band structure of unstrained LiFeAs ("unstrained" refers to the sample without the screw cranked up) is the same as pristine LiFeAs, where two hole-like Fermi surfaces (FSs) are located at the Brillouin zone (BZ) center (Γ) and two electron-like FSs at the BZ corner (M) (Fig. 2A). The uniaxial strain along the [110]-direction (Fig. 2B left) is expected to impact the crystal structure. To apply strain, we built a sample holder which can continuously apply the mechanical pressure or tensile on the mounted sample, with the strength of strain being controlled by rotating the driving screw inside a vacuum chamber (Fig. 2B right). With this design, an *in-situ* strain along one desirable direction can be applied on the sample.

High-resolution ARPES measurements of the band structure in an unstrained LiFeAs sample around Γ with *p*-polarized photons are shown in Figs. 2D and 2E, where the relative position of TI Dirac states can be observed. The band dispersion of LiFeAs under the [110]-direction uniaxial pressure is shown in Figs. 2F and 2G. The comparison of band structures between unstrained and strained samples is shown in Fig. 2C. It can be seen that, under the external strain along the [110] direction, both the bulk bands and the TI surface band shift up in a similar fashion as the hole-doping effect in LiFeAs, and with a strong-enough strain the TI Dirac point can be even tuned above $E_F$ (fig. S1). Besides shifting bands or equivalently $E_F$, the strain effect also changes the FS morphology. In unstrained LiFeAs, the large hole-like FS of the $d_{xy}$ orbital at Γ is square-like as shown in Fig. 2H (39). When the [110]-direction strain is applied, the FS shape of $d_{xy}$ undergoes an obvious deformation to become a $C_2$-symmetry rhombus (Fig. 2I), which reflects the influence of crystal lattice deformation.

**Tunable vortex bound states in LiFeAs**

Inspired by the above APRES results, we carried out a comprehensive STM/S experiment on strained LiFeAs samples and measured their superconducting vortices under the external magnetic field. Note that we specifically use LiFeAs samples from the same batch and the same strain-apply method used in the ARPES measurements (Fig. 2B), to ensure the compatibility between ARPES and STM/S results. We clearly observed the deformation of vortex-core shapes (Fig. 3A), which confirms that the [110]-direction strain has been successfully applied on LiFeAs. As for the unstrained LiFeAs (marked as 'P0' in Fig. 3A), the local density of states (LDOS) distribution around a vortex displays a star-like shape, whose tails are along the As-As directions (33,35), likely due to the quasiparticles scattering along the parallel sides of the $d_{xy}$ FS square (as plotted in the inset of Fig. 2H). After applying the [110]-direction strain in three increasing sequences (marked as 'P1', 'P2' and 'P3' in Fig. 3A), which refers to weakly strained, intermediately strained, and strongly strained, respectively, more details are discussed in the Method.), we observed that the vortex-core shape breaks the original $C_4$-symmetry, and becomes a rectangle shape with the larger length-width ratio under the stronger strain. We point out that this deformation of vortex-core shape is similar to a recent observation on wrinkles of LiFeAs surface, explained as the local strain effect (40). In addition, the deformation of crystal lattice, as indicated by the opposite changes of the lattice lengths along the Fe-Fe directions (fig. S6), also proves the existence and the direction of the external strain.

We next study the strain effect for vortex bound states in LiFeAs under the applied magnetic field. To avoid the unpredictable impurity effect (35, 41), we focus on the impurity-free vortices which are distributed on the homogeneous surface. The STS measurements across an impurity-free vortex in the unstrained sample is plotted in Fig. 3B, where no ZBCP appears and only the dispersive vortex bound states can be seen. As previously discussed, such trivial vortex states are likely produced by the outmost $d_{xy}$ bulk band and the inner $d_{yz}$ bulk band (33,35). We then show the STS measurements across impurity-free vortices in the strained samples (Figs. 3C to 3E). Even with a good anticipation, we still find it remarkable that clear ZBCPs emerge inside most of the impurity-free strained vortices, accompanied by several bound states (the first and second row of Figs. 3C to 3E). Through the waterfall plots (the third row of Figs. 3C to 3E), the intensity of these ZBCPs gradually decreases to zero without splitting when moving away from the vortex center, and pairs of energy-symmetric side peaks display the discrete behavior across the vortex core. These observed in-gap states in the strained vortices are very similar to the topological vortices states referred as MZMs and Caroli-de Gennes-Matricon bound states (CBSs), which have been reported in the other known FeSC Majorana materials (15, 24, 26).

We extract the energy positions of the discrete vortex bound states (marked by $L_0$, $L_{\pm1}$, and $L_{\pm2}$,) by fitting the measured $dI/dV$ spectra (as shown in fig. S5). We find that the energies of these discrete bound states nearly follow the integer level with the ratio of 0: 1: 2 (Fig. 4C). More significantly, their energy spacings ($\Delta E$) can be changed as the applied strain strength is modified (P1, P2, and P3 in Fig. 4C), showing the small-large-small rhythmic response. We also notice that the stronger peak of the first non-zero-energy states ($L_{\pm1}$), as indicated by the black arrows in Fig. 4B, shifts from the positive energy side (P1) to the negative energy side (P3). As explained below, this rhythmic response of vortex bound states corresponds to different positions of the Dirac point when the chemical potential is continuously pushed down by the [110]-direction pressure.

To ensure the reproducibility of our results, we repeated our STM/S measurements on different vortices of several samples. A comparison of a large-scale vortex lattice mapping of the unstrained (Fig. 4D) and strained LiFeAs (Fig. 4E), displaying the nearly uniform deformation of vortex-core shape, suggests the external strain is applied with a relatively high uniformity. In our observation, MZMs can emerge from most of impurity-free vortices under various magnetic fields as long as the [110]-direction strain is applied on LiFeAs within a suitable strength range (fig. S2 and table S1). Statistically, out of 44 impurity-free vortices measured under the strained cases, 33 vortices (75%) have clear MZMs. In contrast, not even one vortex out of 21 impurity-free vortices measured in unstrained samples has MZM. The statistical probability of MZMs increases from 0% to 75%, confirming that a MZM would emerge inside the magnetic vortex of LiFeAs under a suitable strain. We suspect that the reason of 25% vortices without MZMs is due to the nonuniform strained effect when some regions experience insufficient or overloaded strain. Indeed, we observed 100% MZM appearance in the impurity-free vortices under P2 case while the percentage is lower for P1 and P3 cases (Fig. S3), supporting this conjecture. Meanwhile, we observed that the superconducting gaps of LiFeAs are also gradually enlarged with increasing strain (Fig. 4A), indicating the strained sample is still under a robust superconducting region. In addition, the surprising enhancement of superconductivity in LiFeAs under uniaxial strain is opposite to the behavior under hydrostatic pressure or chemical doping (42, 43), which may have important implications to the superconducting mechanism for FeSCs, and will be addressed in a separated paper.

#### Discussion

From the results presented above, we can draw a clear physical picture that MZMs emerge through a vortex phase transition induced by the applied strain in LiFeAs, as schematically displayed in Fig. 4F. When the [110]-direction strain is continually applied on LiFeAs, $E_F$ is gradually pushed downward. Once the applied strain is large enough, $E_F$ will no longer cross the TI surface state twice or leave the “trivial vortex region”(32), and thus MZMs emerge from the topological vortices generated by the superconducting topological surface states (5,13,15,18). Since it is under the quantum limit at the low temperature for this material under our experimental conditions, all the vortex bound states including the MZM are discrete with the nearly integer-level energies (Fig. 4C and fig. S5). In addition, the spatial line profile of the ZBCP can be well fitted by an analytical Majorana wave function (fig. S5I). All these results are the hallmarks of the vortex MZM (20,26).

Furthermore, the energy spacings ($\Delta E$) of the vortex bound states is theoretically proportional to $\Delta^2/E_F$ in the quantum-limit case (20,26). Since different strain strengths generate different values of $E_F$, different energy spacings ($\Delta E$) are expected under different strain strengths. As a result, the energy spacings of P1, P2, and P3 in Fig. 4C show the small-large-small pattern, which is exactly the consequence of the large-small-large absolute values of $E_F$ in these three cases. Therefore, we are able to quantitatively extract the values of $E_F$ for each strained case, as shown in the right column of Fig. 4B, suggesting the continuous downward shift of the chemical potential, fully consistent with our ARPES measurements.

It is also known that the most pronounced first non-zero-energy states ($L_{\pm 1}$) locates at the negative/positive energy side when the Dirac point is above/below $E_F$ (20,26), which explains why the intensity pattern of $L_{\pm 1}$ from P1 to P3 switches side when $E_F$ shifts across the Dirac point (Fig. 4B). As for the trivial vortex states (P0 case), since it is caused by the hole-like parabolic bulk band, the stronger intensity of $L_{-1}$ level is consistent with previous works (33-35). In addition, the spatial decay length of MZMs wave function is inversely proportional to the absolute value of $E_F$ (as simulated by the analytical model in fig. S4E), thus providing a good understanding of the longest decay length of P2 which has the smallest absolute value of $E_F$ (fig. S4D). Thus, we can conclude that the strain-induced evolution of vortex bound states we observed is fully consistent with the expected behavior of the tunable vortex MZM.

This clear picture of how MZM is “squeezed” out from a vortex by uniaxial pressure in LiFeAs, with quantitative “finger prints” of discrete topological vortex states, serves as powerful evidence that the observed zero modes in the strained LiFeAs are indeed MZMs. More importantly, our results suggest that applying external strain is a fine method to control the phase of vortex and to create/annihilate of vortex MZMs in the FeSC system without inducing an inhomogeneous issue. Naturally, the replaced electrical strained device like piezo-stack (44) could rapidly manipulate the procedure of MZMs creation/annihilation, which could in principle satisfy the requirement of recent proposals for non-Abelian Majorana braiding in the Hilbert space without exchanging vortices in the real space (45-47).

**Materials and Methods**

High-quality single crystals of LiFeAs were grown using the self-flux method (*30*). The precursor of Li3As was first synthesized by sintering Li foil and an As lump at ~650 °C for 10 h in a Ti tube filled with argon (Ar) atmosphere. Then the Li3As, Fe and As powders were mixed according to the elemental ratio of LiFe0.3As. The mixture was put into an alumina oxide tube and subsequently sealed in a Nb tube and placed in an evacuated quartz tube. The sample was heated to 1100 °C for 20 h and then slowly cooled down to 750 °C at a rate of 2 °C per hour. Crystals with a size of up to 5 mm were obtained. To protect the samples from reacting with air or water, all the synthesis processes were carried out in a high-purity Ar atmosphere.

High-resolution laser-ARPES measurements on LiFeAs were performed on a spectrometer with a VG-Scienta R4000WAL electron analyser with 6.994-eV photoenergy in the Institute of Solid-State Physics, Japan. The energy resolution of the system was set to ~ 3 meV. The helium-lamp ARPES measurements on LiFeAs were performed on a spectrometer with a VG-Scienta R4000WAL electron analyser with 21.2-eV photon energy in the Institute of Physics, Beijing, China. The energy resolution of the system was set to ~ 6 meV. To obtain clean surfaces for ARPES measurements, the samples was cleaved *in-situ* at ~ 22 K in a cryogenic vacuum.

STM/S measurements were conducted in an ultrahigh vacuum (1 × 10−11 mbar) USM-1300-$^3$He system with a 16-T magnet in the Synergic Extreme Condition User Facility, Beijing, China. The energy resolution is better than 0.26 meV. Tungsten tips were calibrated on a clean Au(111) surface before use. Vertical magnetic fields were applied to the sample surface. All STS data shown in this paper were acquired at 400 mK. STM images were obtained in the constant-current mode. Differential conductance (*dI*/*dV*) spectra and constant bias maps were acquired by a standard lock-in amplitude at a frequency of 973.0 Hz under a modulation voltage $V_{mod}$ = 0.1 mV. All the data acquired by STM/S in this work were measured under the same setpoints: sample bias $V_b$ = −5 mV; tunneling current $I_t$ = 200 *p*A. As the tip-sample separation is sufficiently large, the observed atomic-resolved features in topography correspond to lithium atoms. If the tunnel current is considerably increased in experiments, the atomic resolved topography can show the sites of arsenic atoms (*47*). The zero-bias conductance peaks were reproduced in the impurity-free vortices on three independent samples. To obtain fresh and clean surfaces for STM measurements, the samples were cleaved *in situ* at the room temperature and transferred to the scanner immediately.

# Declaration

**Acknowledgments:** We acknowledge J. R. Huang, B. Jiang, D. Xing, and Y. Dong for experimental assistance, Q. G. Bai for assistance of design and manufacture of the strain device, W.S. Hong and S. L. Li for providing the substrates for strained samples, and W. H. Dong and S. X. Du for useful discussions of the band structure.

**Funding:** The work at IOP is supported by grants from the National Natural Science Foundation of China (11888101, 11234014, 61888102, 11920101005, 11921004), the Ministry of Science and Technology of China (2016YFA0202300, 2018YFA0305700, 2017YFA0302900, 2019YFA0308500), and the Chinese Academy of Sciences (XDB28000000, XDB07000000, 112111KYSB20160061). The work at Japan is supported by JSPS Grants-in-Aid for Scientific Research (JP19H01818, JP19H05826, JP21H04439), and MEXT as the "Program for Promoting Researches on the Supercomputer Fugaku" (JPMXP1020200104). This work is partially supported by the Synergic Extreme Condition User Facility, Beijing, China.

**Author Contributions:** W.L. and H.D. designed the experiments. Q.H. and W.L. performed STM experiments with assistance of F.Y., Y.Z. and W.L. performed ARPES experiments. X.W., Y.P. and C.J. synthesized samples. W.L., Y.Z., and Q.H. analyzed the data with inputs from all other authors. All the authors participated experiment setup and discussions. W.L. and H.D. wrote the manuscript with input from all other authors. H.D., H.G. and J.X. supervised this project.

**Competing Interest :** The authors declare that they have no competing interests.

**Availability of data and material :** All data generated or analyzed during this study are included in this article and its supplementary information files.

**Figures and Tables**

**Fig. 1. Topological band structure and vortex bound states of pristine LiFeAs.** (**A**) Crystal structure of LiFeAs. (**B**) Atomically-resolved STM topography of the Li-terminating surface after cleavage. The white axes indicate the Fe–Fe bond directions. Insert: the lattice viewed along the c-axis of the crystal. (**C**) First-principles calculations of the band structure of LiFeAs [Adopted from Fig. 1f of Zhang *et al* (*21*)], the red circle indicates the position of topological bands. (**D**) Schematic depiction of the topological band structure in LiFeAs. The lower Dirac surface band and the upper bulk TDS bands are observed from laser-ARPES measurements (*21*). When $E_F$ is located at the bending region (yellow shaded region) or slightly above TI surface state, the phase of vortex is trivial (**E**) Typical tunneling conductance spectra are measured at the clean area of a pristine LiFeAs sample ($V_s$ = -5 mV, $I_t$ = 200 pA, $T_{exp}$ = 0.4 K). The 3 T - *dI*/*dV* spectrum (red) is measured at the center of this vortex which has a sharp non-zero-energy vortex bound, and the zero-field spectrum (black) is measured at the same position, displaying two bulk superconducting gaps: $\Delta_1$ = 2.9 meV, and $\Delta_2$ = 5.8 meV. Insert: the zero-bias conductance (ZBC) map of the same vortex. **(F) and (G)** Sketches of the ordinary vortex case in the pristine LiFeAs, and the topological vortex case with MZMs in the Fermi-level-tuned LiFeAs. When the pristine LiFeAs is at the topological trivial phase, only two MZMs annihilate and no ZBCP is observed. However, once the chemical potential of LiFeAs is tuned to the topological nontrivial phase (indicates by the blue shaded region of (D)), MZMs are expected to be observed inside vortices.

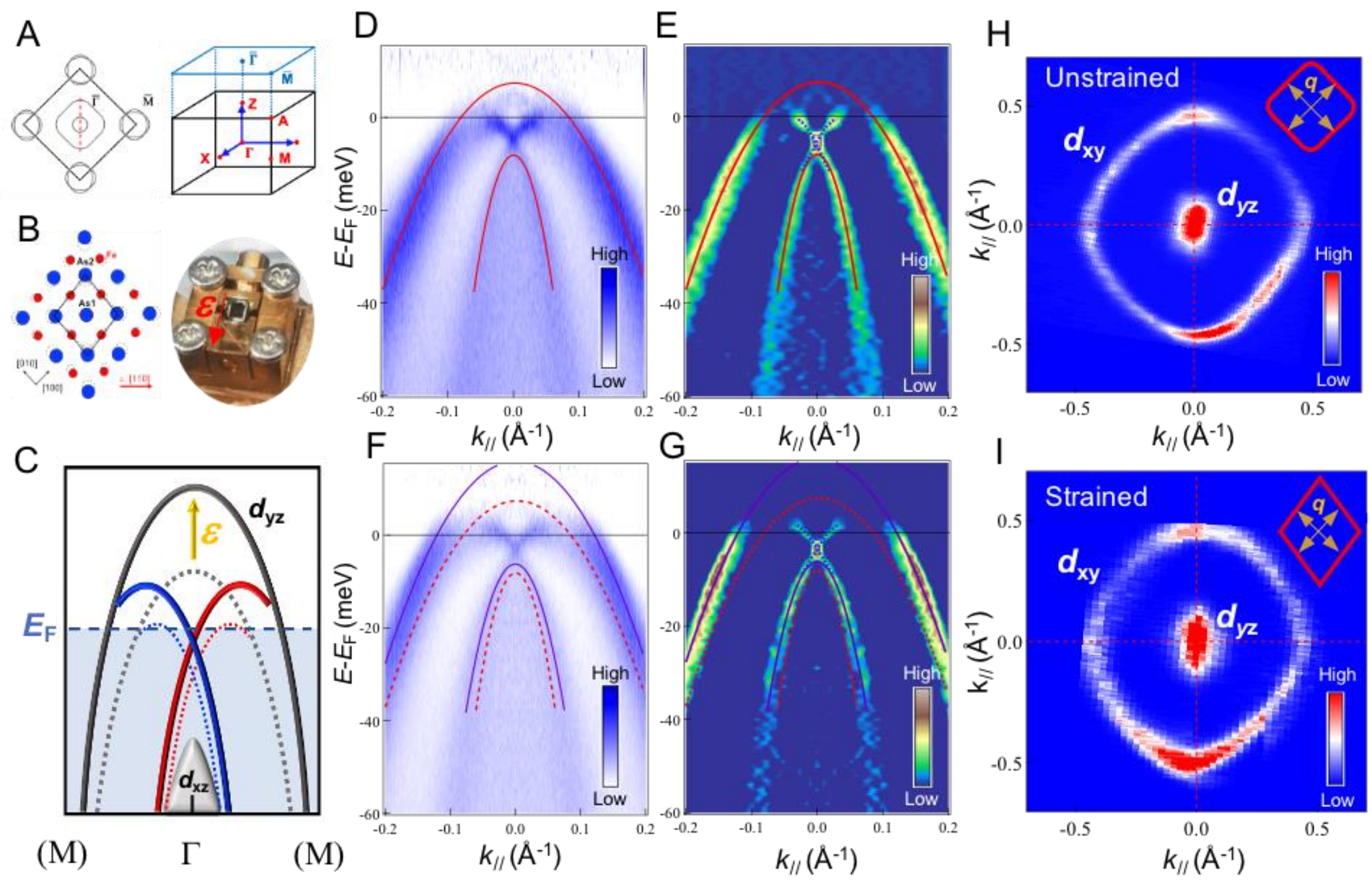

**Fig. 2. Strain-induced band shift in LiFeAs.** (**A**) Schematic of surface BZ, and the projection of 3D BZ on the (001) surface in LiFeAs. There are two hole-like FSs at $\Gamma$ and two electron-like FSs at M. (**B**) Left panel: transformation of LiFeAs crystal lattice under the [110]-direction uniaxial compressive strain $\varepsilon$. Dashed open circles represent the atomic positions in the unstrained unit cell. Right panel: picture of the *in-situ* strain-tuning sample holder (with sample), a red arrow indicates the applying direction of strain. Strain can be sequentially enhanced by rotating the horizontal screw. (**C**) Strain effect on the band structure of LiFeAs, where both the bulk bands and the Dirac surface band shift upward under the [110]-direction uniaxial compressive strain. (**D**) ARPES intensity plot of the unstrained LiFeAs along $\Gamma$-M at 24 K, with a laser delivering *p*-polarized 7-eV photons. The 2D spectrum is divided by the corresponding Fermi function. The red lines are parabolic-type fitting of the MDC peaks, which extracts the band dispersion. (**E**) MDC second-derivative plot of (D). **(F) and (G)** are the same as (D) and (E) but on the strained LiFeAs. The purple solid lines extract the band dispersion of the strained LiFeAs and the red dashed lines represent the unstrained case. The blue dashed lines in (E) and (G) are the guides to the eye, indicating the Dirac dispersion. **(H) and (I)** Fermi surfaces measured by Helium lamp ($h\upsilon$ = 21.2 eV) on the unstrained and strained LiFeAs, respectively. The shape of the $d_{xy}$ FS pocket is deformed from square to rhombic under strain. Insert: vector *q* indicates the traveling direction of quasiparticles on the $d_{xy}$ FS.

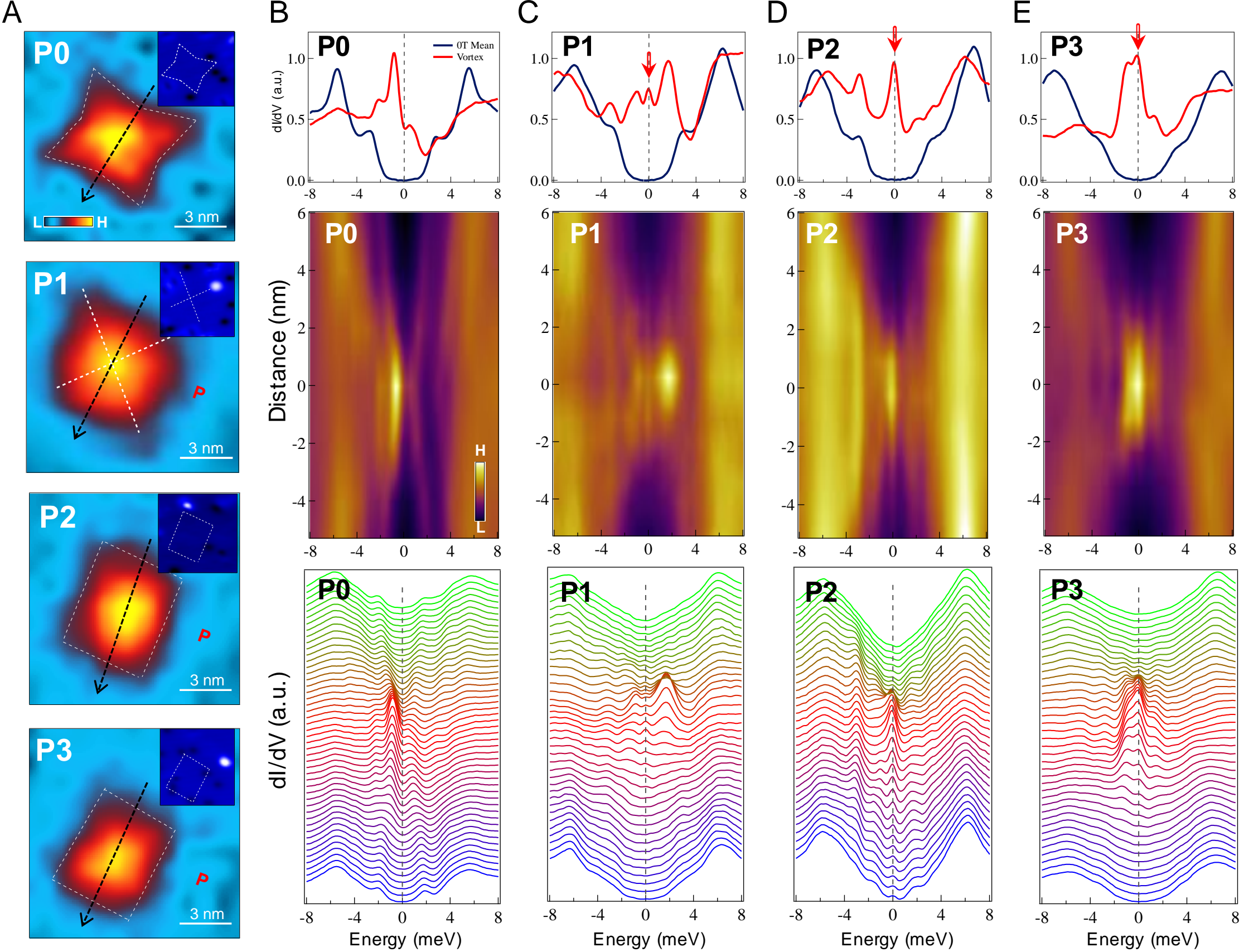

**Fig. 3. Vortex evolution with strain in LiFeAs. (A)** ZBC maps around vortices under different [110]-direction strain strength marked by 'P0' to 'P3': P0 vortex is on the unstrained LiFeAs, and P1 - P3 cases refer to a sequence from weak strain (P1) to strong strain (P3). P0 vortex exhibits a star-like shape with tails along the As-As direction. With increased strain, the shape of vortices is deformed. The red arrows with the mark 'P' indicate the direction of the strain. Inset: corresponding STM topography, the white dashed boxes indicate the locations of vortices. (**B**) *dI/dV* result of P0 vortex: the upper panel shows *dI/dV* spectra near the vortex center and the spatial-averaged 0-T spectra, the red arrows highlight the existence of ZBCPs; the middle panel is the line-cut intensity plot measured under a magnetic field; the lower panel is the waterfall-like plot of the middle panel. **(C) – (E)** are the same as (B) but for P1, P2, P3 vortex cases, respectively. P0 - P2 vortices are measured under 4 T, and P3 vortex is measured under 5 T. All line-cuts are measured along with the according black arrow dashed lines indicated in (A). With the strain applied on the LiFeAs, the clear ZBCPs appear accompanying the discrete in-gap bound states.

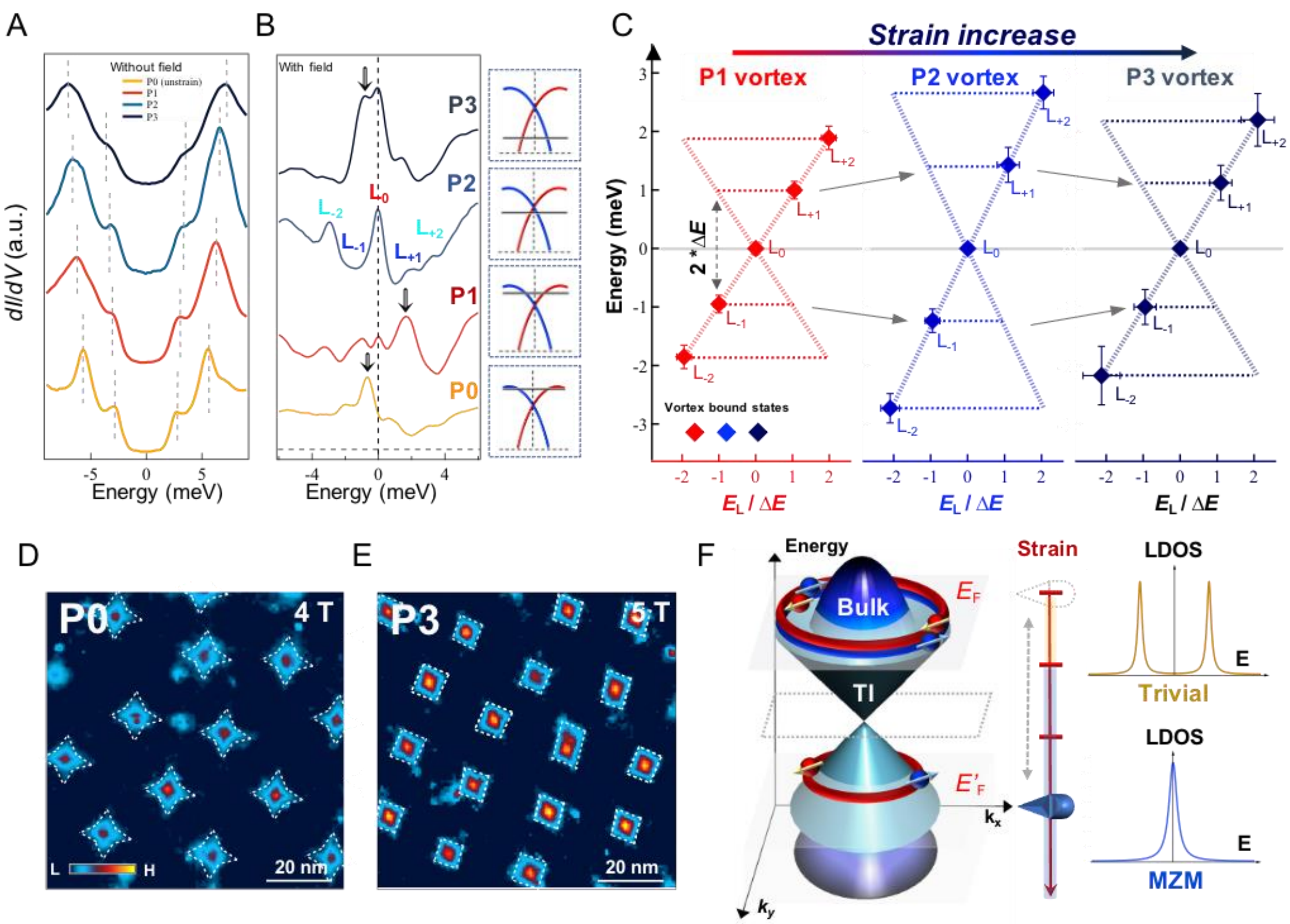

**Fig. 4. Tunable vortex MZMs in LiFeAs. (A)** Comparison of 0-T *dI/dV* spectra from Figs. 3 (B) – (E). The superconducting gaps are enlarged with increasing strain, e.g., for P0: $\Delta_1 =$ 2.9 meV, and $\Delta_2 =$ 5.8 meV, and for P3: $\Delta_1 =$ 3.7 meV, and $\Delta_2 =$ 7.0 meV, with $\Delta_1$ ($\Delta_2$) attributed to the outer (inner), or $d_{xy}$ ($d_{yz}$) FS (*45*). **(B)** Comparison of *dI/dV* spectra near vortex centers from Figs. 3 (B) – (E). The colored markers of '$L_0$', '$L_{\pm1}$', and '$L_{\pm2}$' represent the discrete in-gap bound states at different energies ($E_L$), which are obtained by the fittings shown in fig. S4C. Black arrows indicate the locations of the stronger one of the $L_{\pm1}$ bound states. The approximate locations of $E_F$ of the different strain cases (P0 to P3) are draw in the sketches on the right. **(C)** Summary of the energies of $E_L$ and ratios of $E_L/\Delta E$ of vortex bound states in strained vortices. $\Delta E_1 = 0.95$ meV, $\Delta E_2 = 1.3$ meV, $\Delta E_3 = 1.0$ meV, and $E_F$ can be estimated for each case via the formular of $\Delta E = (\Delta_{sc})^2/E_F$. and all the ratios of $E_L/\Delta E$ are nearly integers. To improve the reliability, each labeled data point is extracted as the statistical average of all the measured impurity-free-vortex bound states under each strained case, *i.e.,* P1, P2, and P3 cases. The error bars are the root-mean-square values of the vortex-bound-states statistics. The involved strained vortices are partly shown in fig. S2, and the detailed fitting procedure and analysis are shown in fig. S4. **(D) and (E)** ZBC maps (area: 90 nm × 90 nm) of the strained and unstrained LiFeAs under a magnetic field, confirming uniform deformation of vortices. **(F)** Vortex phase transition in LiFeAs. Under the uniaxial strain, the chemical potential is *in-situ* tunable, and thus

the trivial vortex case can be transformed to the topological vortex case which contains the isolated MZM well separated from other discrete bound states.